\documentclass[sn-mathphys,Numbered]{sn-jnl}% Math and Physical Sciences Reference Style
\usepackage{makecell}
\usepackage{graphicx}%
\usepackage{multirow}%
\usepackage{amsmath,amssymb,amsfonts}%
\usepackage{amsthm}%
\usepackage{mathrsfs}%
\usepackage[title]{appendix}%
\usepackage{xcolor}%
\usepackage{bm}%
\usepackage{textcomp}%
\usepackage{manyfoot}%
\usepackage{booktabs}%
\usepackage{algorithm}%
\usepackage{algorithmicx}%
\usepackage{algpseudocode}%
\usepackage{listings}%
\theoremstyle{thmstyleone}%
\newtheorem{theorem}{Theorem}%  meant for continuous numbers
\theoremstyle{thmstyletwo}%

\theoremstyle{thmstylethree}%

\DeclareMathOperator{\sgn}{sgn}
\DeclareMathOperator{\grad}{\bm\nabla}

\newcommand\highlightReference[1]{%
  \expandafter\newcommand\csname highlightReference-#1\endcsname{}%
}
\let\oldbibitem\bibitem
\def\bibitem#1 #2\par{%
  \expandafter\ifx\csname highlightReference-#1\endcsname\relax
    \oldbibitem{#1}#2
  \else
    \oldbibitem{#1}\highlight{#2}
  \fi
}
\usepackage{color}
\newcommand\highlight[1]{{#1\hspace{-1pt}}}

\begin{document}

\title[Regularised Diffusion--Shock Inpainting]
      {Regularised Diffusion--Shock Inpainting}


\author*[1]{\fnm{Kristina} \sur{Schaefer}}\email{schaefer@mia.uni-saarland.de}

\author*[1]{\fnm{Joachim} \sur{Weickert}}\email{weickert@mia.uni-saarland.de}

\affil*[1]{\orgdiv{Mathematical Image Analysis Group, Dept. of Mathematics and Computer Science}, 
\orgname{Saarland University}, \orgaddress{\street{E1.7}, \postcode{66041}
\city{Saarbr\"ucken},  \country{Germany}}}

\abstract{
We introduce regularised diffusion--shock (RDS) inpainting as a modification 
of diffusion--shock inpainting from our SSVM 2023 conference paper. 
RDS inpainting combines two carefully chosen 
components: homogeneous diffusion and coherence-enhancing shock filtering.
It benefits from the complementary synergy of its building blocks:
The shock term propagates edge data with perfect sharpness 
and directional accuracy over large distances due to its high degree of 
anisotropy.
Homogeneous diffusion fills large areas efficiently. The second order equation 
underlying RDS inpainting inherits a maximum--minimum principle from its 
components, which is also fulfilled in the discrete case, in contrast to 
competing anisotropic methods. The regularisation addresses the largest 
drawback of the original model: 
It allows a drastic reduction in model parameters without any loss in quality. 
Furthermore, we extend RDS inpainting to vector-valued data. 
Our experiments show a performance that is comparable to or better than many 
{ inpainting methods based on partial differential equations and 
related integrodifferential models}, including anisotropic processes of second 
or fourth order.

}

\keywords{shock filters, inpainting, diffusion, 
    mathematical morphology, image processing}



\maketitle

%------------------------------------------------------------------------------
\section{Introduction}
Image inpainting \cite{EL99a,MM98a} is the task of filling in missing regions 
in an image. There are many approaches for solving this task, but in this
work we focus on inpainting based on partial differential equations (PDEs).
This class of methods is particularly successful in applications with very 
sparse data such as image compression~\cite{GWWB08,JPW20,SPME14}. 

PDE-based inpainting methods are often inspired by physical processes. 
For instance homogeneous diffusion {\cite{Ii62,Ii63a,WII97}} 
is inspired by heat propagation, and Euler's elastica 
inpainting \cite{MM98a,Mu94a} is connected to the elasticity of solids.

Creating a high quality inpainting result with PDE-based methods has some
particular challenges. Many operators struggle to bridge large gaps,
introduce dissipativity into high contrast images (such as binary ones), or
do not reproduce the direction of structures accurately. 
It is often assumed that high order PDEs such as Euler's 
elastica \cite{MM98a,Mu94a} or Cahn--Hilliard inpainting \cite{BEG07} 
are necessary to address these challenges. However,  edge-enhancing 
diffusion (EED) \cite{We94e} as a second order integrodifferential process
has been 
shown to provide the desired properties in practice as well~\cite{SPME14}. 

One useful property in the context of inpainting is the fulfilment of a
maximum--minimum principle which guarantees that no over- and undershoot
are introduced. Most higher order methods violate this principle.
EED satisfies a maximum--minimum principle in the continuous case, but
to date there is no discretisation with reasonably small 
stencils available that inherits this property.

\medskip

\textbf{Contributions.}
In order to address these challenges, we have proposed  diffusion--shock 
inpainting in our conference publication \cite{SW23}. It is a PDE-based 
inpainting operator that fulfils the desired properties in practice, while 
also providing a maximum--minimum principle in the discrete case. This 
is achieved by combining two time-proven methods: homogeneous diffusion 
{\cite{Ii62,Ii63a,WII97}} and coherence-enhancing shock 
filtering \cite{We03}. 
Originally designed with the goal of deblurring, shock filters create sharp 
edges at the boundary between influence zones of maxima and minima 
{ by using the sign of a second derivative 
operator}~\cite{KB75,OR90}. 
However, the coherence-enhancing shock filter can also propagate image 
structures over large distances without directional or dissipative 
artefacts, which can be seen in Fig.~\ref{fig:line}(c). In diffusion--shock
inpainting, the shock filter propagates edges of image structures without 
introducing dissipative artefacts. From the newly created structures 
the homogeneous diffusion term fills in larger homogeneous areas. 
The synergy of these two methods allows high quality results. Even for 
high contrast images it reconstructs edges with
perfect sharpness and high directional accuracy. Our numerical algorithm
satisfies a maximum--minimum principle, and it is optimised for rotation
invariance. The experiments in \cite{SW23} show that diffusion--shock 
inpainting produces results
that rival the quality  of state-of-the-art PDE-based inpainting methods 
such as EED and Euler's elastica.

\medskip

In addition to our original conference publication \cite{SW23}, we make 
the following novel contributions in this work:
\begin{enumerate}
\item We introduce regularised diffusion--shock (RDS) inpainting as a
  regularised version of diffusion--shock inpainting. 
  { To this end we replace the $\sgn$ function that acted as
  guidance for our original diffusion--shock inpainting model by a 
  sigmoid-like function.}
  This stabilises the process w.r.t. the parameter choice, which 
  allows us to establish a parameter coupling without loss of quality. 
  Thereby we reduce the number of parameters to two, which 
  makes the model more accessible in practice. 
\item We give a more detailed description of the numerics. 
\item We compare the performance of RDS inpainting with many 
  related approaches. This systematic evaluation reveals that the
  coherence-enhancing shock term is crucial to the success of RDS inpainting. 
\item Finally, we extend our model to vector-valued data which allows the
  application to colour images. 
\end{enumerate}

\medskip
\textbf{Related Work.} 
With the goal of image deblurring, Kramer and Bruckner \cite{KB75} have
proposed a first discrete model of a shock filter already in 1975. Later
Osher and Rudin \cite{OR90} have formulated a first PDE-based approach and 
coined the term shock filter.  Shock filters typically utilise a second 
derivative operator to identify the influence zones of maxima and minima. 
Osher and Rudin \cite{OR90} have considered the Laplacian  as well as the  
second derivative in gradient direction. Alvarez and Mazorra \cite{AM94} have
introduced presmoothing to the second derivative operator in order to 
robustify the process against noise. {As another strategy, Diop and 
Angulo \cite{DA20} propose to locally adapt the shock filter to the image 
to reduce the sensitivity to noise.}
The  coherence-enhancing shock filter of Weickert \cite{We03} relies on 
the second directional derivative in the dominant eigendirection of the 
structure tensor \cite{FG87}. 
In the next section, we will cover {the shock filters that are 
relevant for this paper in more detail.}
While theoretical results for continuous shock filters are rare, Welk et 
al.~\cite{WWG07} have established well-posedness of 1-D space-discrete and 
fully discrete shock filters.

Inspired by the implicit presence of shock terms within nonlinear evolutions 
such as Perona--Malik diffusion \cite{PM90}, self--snakes~\cite{Sa96} or
the PDE-based version of the Kuwahara--Nagao operator \cite{Bo02}, many 
explicit combinations have been proposed. Typically the shock term of Alvarez 
and Mazorra \cite{AM94} is combined with homogeneous 
diffusion~{\cite{Ii62,Ii63a,WII97}}, e.g~\cite{KDA97,FRWC06}, 
or mean curvature 
motion~\cite{Br78}, e.g \cite{AM94,KDA97,XPZKYW16}. Gilboa et al. \cite{GSZ02}
rely on complex diffusion. Usually 
these combinations are used in the context of image enhancement, but not 
in image inpainting.

RDS inpainting is one of the rare examples of hyperbolic PDEs in inpainting.
Another exception is the method of Bornemann and M\"arz \cite{BM07},
{ which was extended by M\"arz in \cite{M11}}. It relies on 
transport processes that are guided by structure tensor information. Therefore,
 it is close in spirit to RDS inpainting. However, their paper 
follows a more algorithmic approach without specifying a compact 
evolution equation. In our experiments we will compare against this method.
 Another approach that relies on a hyperbolic concept is  the recent 
inpainting model of Novak and Reini\'c \cite{NR22}. It combines 
a shock filter with the fourth order Cahn--Hilliard PDE. RDS inpaiting 
is conceptually simpler, as it already achieves the desired filling-in effect
with a second order homogeneous diffusion PDE.

To evaluate the performance of RDS inpainting, we compare it 
to various other PDE-based inpainting operators in our experiments. 
This includes  linear and isotropic processes such as homogeneous 
diffusion~{\cite{Ii62,Ii63a,WII97,Ca88}} and  biharmonic 
interpolation \cite{Du76} as its fourth order counterpart.
We also consider nonlinear isotropic processes such as total variation 
(TV) inpainting \cite{SC02}, which 
can be interpreted as a limiting case of Perona--Malik \cite{PM90} 
inpainting with a scalar-valued Charbonnier diffusivity \cite{CBAB97}.
Moreover, we compare our model to anisotropic approaches such as 
Tschumperl\'e's model \cite{Ts06}, which relies on a tensor-driven 
equation that uses the curvature of integral curves, and to
edge-enhancing diffusion \cite{We94e}, which is the core of the 
state-of-the-art-image compression codec R-EED \cite{SPME14}. Furthermore, 
we consider the popular higher order inpainting method based on 
Euler's elastica \cite{MM98a,Mu94a}.

{
Deep learning techniques have gained popularity for solving inpainting 
tasks in the past decade~\cite{PKDD16,YLYS19}. Especially the recent 
approaches based on probabilistic diffusion~\cite{SDWMG15,HJA20,RBLEO22} have
sparked a public discussion due to their highly realistic image generation 
capabilities. While such models can work well in practice, they typically 
involve a huge number of parameters that make it very difficult to gain 
a deeper understanding of their inner workings. Furthermore, they usually 
do not provide any formal guarantees.
On the other hand, our RDS inpainting is a PDE-based model that relies on
time-proven components that are carefully selected for the task of image 
inpainting. The corresponding numerics relies on schemes that are
well understood, and it satisfies a maximum--minimum principle.
Comparing these two opposite ideologies would not do justice to either of them.
Therefore, we do not compare our method to purely learning-based approaches. 
However, neural networks may also incorporate model-based ideas \cite{GFE21}.
This can be used for the implementation of numerically challenging models;
e.g. \cite{SAWE22} uses a neural network for solving Euler's elastica for 
image inpainting. 
In our experiments, we compare our RDS inpainting to this hybrid approach.
}

\medskip
\textbf{Organisation of the Paper.} In Section \ref{sec:reviewshock}, 
we review the concept
of shock filters. Section \ref{sec:DS} introduces the RDS inpainting
model in the continuous setting. A numerical scheme with high rotation 
invariance and stability guarantees in the maximum norm is discussed in Section 
\ref{sec:discretisation}. We evaluate our model experimentally in Section 
\ref{sec:experiment}, before concluding the paper in Section 
\ref{sec:conclusions}.

%%%%%%%%%%%%%%%%%%%%%%%%%%%%%%%%%%%%%%%%%%%%%%%%%%%%%%%%%%%%%%%%%%%%%%%%%%%%%%%
\section{Review of Shock Filters}
\label{sec:reviewshock}
Shock filters have been introduced with the goal of image sharpening and 
deblurring. By propagating the values of extrema to their influence zones, 
shocks are formed at the boundary of these zones. The various ways of 
characterising these influence zones create
 different shock filter models that we briefly review in this 
section. 

%------------------------------------------------------------------------------
\subsection{PDE-based Morphology}
For brightening and darkening of image regions, shock filters rely on the 
building blocks of mathematical morphology~\cite{So04}: dilation and erosion. 
The dilation $\oplus$ of a grey value image 
$f:\Omega\subset\mathbb{R}^2\to\mathbb{R}$ replaces the image value in a 
location $\bm x$ by its supremum within a neighbourhood $B$, the so-called
structuring element.\footnote{ Throughout our paper
vectors are denoted by lower case boldface letters and matrices by upper 
case boldface letters.} The erosion 
$\ominus$ uses the infimum instead. The operations are defined as
\begin{align}
(f\oplus B)(\bm x) \; &= \; \sup \{ f(\bm x - \bm y) \, | \, \bm y \in B\},\\
(f\ominus B)(\bm x)\; &=\; \inf\;\{ f(\bm x + \bm y)\, | \, \bm y \in B\}.
\end{align}
For shock filters, their PDE-based formulations are more 
popular. Dilation/erosion $u$ with a disk-shaped neighbourhood of radius $t$
correspond to the solution $u(\bm{x},t)$ of 
\begin{equation}
\partial_t u \; = \; \pm\, |\bm\nabla u|
\end{equation}
with the initial image $u(\bm x, 0) = f(x)$ and reflecting boundaries 
\cite{BM92,AGLM93,AVK93}. 
The $+$ sign corresponds to dilation, and $-$ yields erosion. We denote the 
spatial nabla operator by $\bm\nabla = (\partial_x, \partial_y)^\top $, and 
$|\cdot|$ is the Euclidean norm. 

%------------------------------------------------------------------------------
\subsection{Shock Filters}
In order to achieve the desired sharpening, shock filters apply 
dilation and erosion adaptively: In influence zones of maxima they use 
dilation, and in influences zones of minima they apply erosion. This switch
is modelled by considering the sign of a second derivative operator. In 
general, shock filters have the form 
\begin{equation}
\partial_t u \; = \; - F(Lu) |\bm\nabla u|\, .
\end{equation}
The \emph{guidance term} $F(Lu)$ determines the behaviour of the shock filter.
It consists of the second order derivative operator $Lu$ and the 
\emph{guidance function} $F:\mathbb{R}\to[-1,1]$, which has to retain the 
sign of its input.

\medskip

\begin{figure}[tb]
\centering
\begin{tabular}{ccc}
\includegraphics[width=0.25\textwidth]{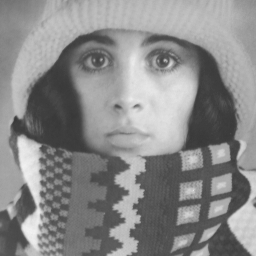} &
\includegraphics[width=0.25\textwidth]{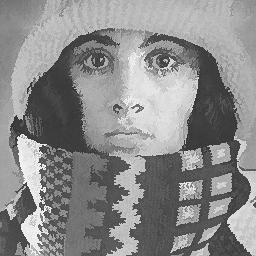}  &
\includegraphics[width=0.25\textwidth]{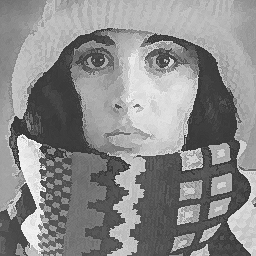} \\
\footnotesize{(a) input} & \footnotesize{(b) $\Delta u$} 
& \footnotesize{(c) $\partial_{\bm\eta\bm\eta} u$}
\end{tabular}
\caption
{Visual comparison of the steady states of shock filters with different
guidance terms with $F(u)= \sgn(u)$.}
\label{fig:shock}
\end{figure}

We distinguish the shock filter types by their second order 
derivative operator $Lu$. Osher and Rudin~\cite{OR90} considered the Laplacian 
$Lu =\Delta u$ and the second derivative $Lu=\partial_{\bm{\eta \eta}} u$ in 
the normalised gradient direction $\bm{\eta} \parallel \bm{\nabla} u$.
They argue that $\partial_{\bm\eta\bm\eta} u$ gives better results. 
This is in accordance with findings of Haralick~\cite{Ha84}, who favours the 
zero crossings of $\partial_{\bm\eta\bm\eta} u$ over the ones of $\Delta u$ 
as edge detectors.
We confirm
this in Fig. \ref{fig:shock}. Both filters result in a non-flat, 
segmentation-like steady state and sharpen the image without drastically 
changing its structure. However, the second derivative in gradient direction 
yields cleaner edges.

\medskip

To robustify the process against noise, Alvarez and Mazorra \cite{AM94} 
introduced a presmoothing to the derivative operator and used 
$Lu = \partial_{\bm\eta\bm\eta} u_\sigma$, where 
{$u_\sigma = K_\sigma * u$} 
denotes the convolution of the image with a Gaussian of standard deviation 
$\sigma$. Applying this presmoothing may drastically change the structure of 
the evolving image.

\medskip

For his coherence-enhancing shock filter, Weickert \cite{We03} uses the 
second derivative in direction of the dominant eigenvector $\bm w $ 
(i.e. the eigenvector with the larger corresponding eigenvalue) of the 
structure tensor 
$\bm J_\rho(\bm\nabla u) = K_\rho *(\bm\nabla u \bm\nabla u^\top)$ 
\cite{FG87}.  Hence, the coherence-enhancing shock filter relies on  
$Lu = \partial_{\bm w\bm w} u_\sigma$. 
We use $\bm J_\rho(\bm\nabla u_\sigma)$ instead of 
$\bm J_\rho(\bm\nabla u)$ since it yields better results for our application.
{
Hence, the dominant eigenvector $\bm w$ depends on the noise scale $\sigma$
and the integration scale $\rho$. As is common in the structure tensor 
literature, we do not make this explicit by adding extra indices.}
As the dominant eigenvector of the structure tensor corresponds to the 
direction of the largest local contrast, this filter has a coherence-enhancing
effect. Similar observations exist in the context of coherence-enhancing 
diffusion \cite{We97d}. 

\medskip

While the choice of $Lu$ determines the main behaviour of the shock filter,
one may also choose from various guidance functions $F$. In our conference 
publication~\cite{SW23} we relied on the $\sgn$ function as the most widely 
used choice. However, different $\mbox{sigmoid}$-like functions have been 
used in the literature, including $\arctan $ functions~\cite{GSZ02} or 
hyperbolic tangent functions~\cite{FRWC06}. For our RDS inpainting, we rely 
on $\mbox{sigmoid}$-like functions as a regularised  alternative to the 
$\sgn$ function. We will evaluate the benefits of this choice in our 
experiments.

\medskip

\begin{figure}[tb]
\centering
\setlength{\fboxsep}{1pt}
\begin{tabular}{ccc}
\fbox{\includegraphics[width=0.25\textwidth]{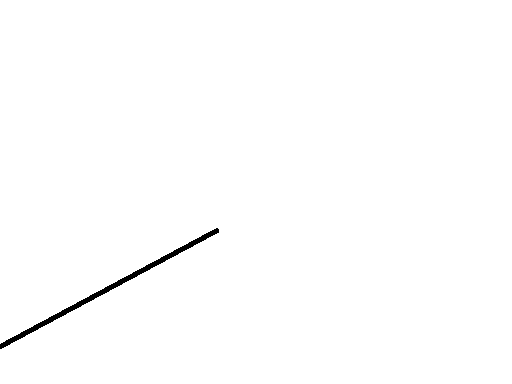}}&
\fbox{\includegraphics[width=0.25\textwidth]{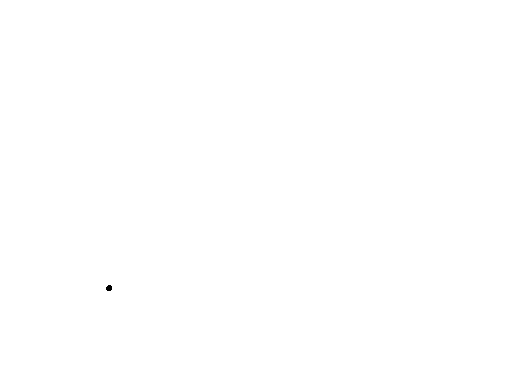}}&
\fbox{\includegraphics[width=0.25\textwidth]{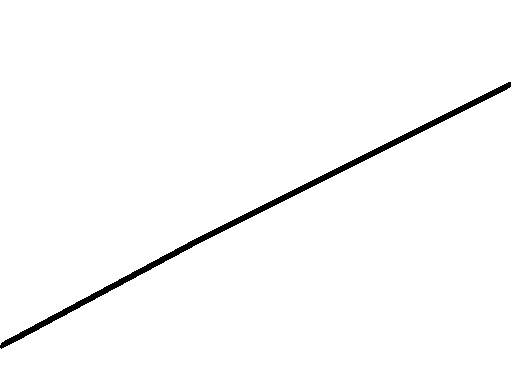}}\\
\footnotesize{(a) input} &  
 \footnotesize{(b) $Lu = \partial_{\bm\eta\bm\eta} u_\sigma$}
& \footnotesize{(c) $Lu = \partial_{\bm w\bm w} u_\sigma$}
\end{tabular}
\caption{Alteration of an edge-like structure by different shock filters with 
presmoothing. 
The steady states 
of the different shock filters with $F(u)= \sgn(u)$, $\sigma = 2$ and 
$\rho = 5$ are shown.}
\label{fig:line}
\end{figure}

In Fig. \ref{fig:line}, we investigate the potential of shock filters
to propagate structures over large distances by the example of a partial line. 
The Alvarez--Mazorra model shrinks the line to a small 
disk-like shape. The coherence-enhancing shock filter elongates the 
line perfectly over a distance of more than 200 pixels in a direction that is 
not grid aligned. Moreover, it creates a perfectly sharp result without 
introducing any dissipativity. This quality is exceptional for PDE-based 
methods. Therefore, we choose the coherence-enhancing shock filter as a key 
component of our RDS inpainting.  

%..............................................................................

%..............................................................................

%%%%%%%%%%%%%%%%%%%%%%%%%%%%%%%%%%%%%%%%%%%%%%%%%%%%%%%%%%%%%%%%%%%%%%%%%%%%%%%
\section{Regularised Diffusion--Shock Inpainting}
\label{sec:DS}
For image inpainting, we decompose the rectangular image domain 
$\Omega $ into two regions: The known data locations
 are represented by the \emph{inpainting mask} $K\subset\Omega$, and
the unknown values are located in the \emph{inpainting domain} 
$\Omega \setminus K$. In the inpainting domain, a PDE-based 
inpainting method applies a suitable differential operator  until the process 
converges. For RDS inpainting a weighted combination of a regularised
coherence-enhancing shock filter and homogeneous diffusion takes that role.

As we show in Fig. \ref{fig:line}, the
coherence-enhancing shock filter can propagate edge-like structures over 
arbitrarily large distances with perfect sharpness and directional accuracy. 
However, the width of the created structures is limited by the 
presmoothing scale $\sigma$. 
Here, homogeneous diffusion is the ideal partner: It efficiently fills the
missing large areas from the sharp edges created by the shock filter. 

In order to achieve this behaviour, we apply a weighted combination of the 
two components such that the shock term dominates near edges, and the 
diffusion term takes over in more homogeneous regions. We model this by
means of a Charbonnier weight function \cite{CBAB97} 
\begin{equation}
  g\left(|\bm \nabla u_\nu|^2\right) \; = \; 
  \frac{1}{\sqrt{1+ |\bm \nabla u_\nu|^2/\lambda^2}}
\end{equation}
with the Gaussian-smoothed image $u_\nu = K_\nu * u$.
It is a decreasing function with range $(0, 1]$, for which we have $g(0) = 1$ 
and   $g(|\bm \nabla u_\nu|^2) \to 0$ for $|\bm \nabla u_\nu|^2 \to \infty$.
By presmoothing the image before computing the gradient, we locally average 
structural information and stabilise the process w.r.t. noise. 

\begin{figure}
\centering
\includegraphics[width=0.5\textwidth]{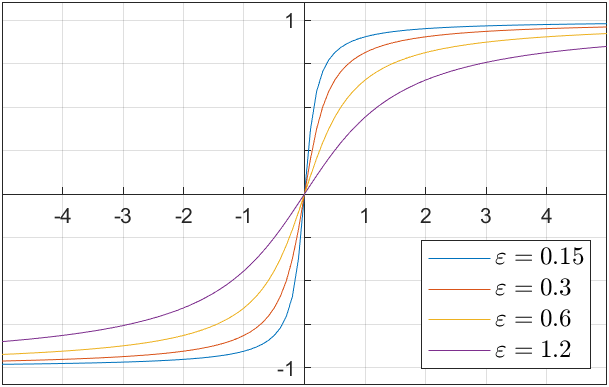}
\caption{Effect of the regularisation parameter $\varepsilon$ 
on $S_\varepsilon(x)=\frac{2}{\pi}\arctan\left(\frac{x}{\varepsilon}\right)$.}
\label{fig:atan}
\end{figure}

With that{,} our \emph{regularised diffusion--shock (RDS) inpainting} 
is based on the PDE 
\begin{equation}
\label{eq:inpainting}
\partial_t u \; = \; g\left(|\bm \nabla u_\nu|^2\right)\, \Delta u
  \;-\; \Big(1-g\left(| \bm \nabla u_\nu |^2\right)\!\Big) \,
  S_\varepsilon\left( \partial_{\bm w\bm w} (u_\sigma) \right)\,
  | \bm \nabla u | \, .
\end{equation}
We use Dirichlet data at the boundaries $\partial K$ 
of the inpainting mask and reflecting boundary conditions on the 
image domain boundary $\partial \Omega$. 
By $S_\varepsilon$ we denote a sigmoidal function with a regularisation 
parameter $\varepsilon>0$. This adds additional regularisation to the model: 
It softens the transition from dilation to erosion in the shock term. 
This  choice is reminiscent of the regularisation of the
Chan--Vese model for segmentation \cite{CV01a}, which relies on a rescaled 
family of $\arctan$ functions. In our experiments, we will use 
\begin{equation}
S_\varepsilon(x) \;=\; \frac{2}{\pi} 
\arctan\left( \frac{x}{\varepsilon}\right) \, .
\end{equation}
As depicted in Fig. \ref{fig:atan} the regularisation parameter $\varepsilon$ 
determines the steepness of the $\arctan$ function. For $\varepsilon\to 0$, 
we arrive at the 
diffusion--shock inpainting model from our conference publication \cite{SW23},
which uses a $\sgn$ function instead. 

%..............................................................................
%------------------------------------------------------------------------------
\subsection{Parameter Coupling}
The model parameters of RDS inpainting fall into two natural categories:
The noise scale $\sigma$, the integration scale $\rho$, and the edge scale 
$\nu$ serve as spatial scale parameters within the image domain,
whereas the contrast parameter $\lambda$ and the regularisation parameter 
$\varepsilon$ are tonal scale parameters acting in the codomain.
This classification allows to introduce a parameter coupling that reduces the 
five parameters to only two. This greatly eases the practical applicability of 
RDS inpainting.

\medskip

The noise scale $\sigma$ determines the width of the structures 
created by the shock term. In the computation of the structure tensor 
$\bm J_\rho(\bm\nabla u_\sigma)$ it removes noise and small-scale details. 
In order to  avoid cancellation effects of gradients with opposite orientation
and very wide borders of edge-like structures $\sigma$ should be chosen 
relatively small. The integration scale $\rho$ allows averaging of directional 
information without cancellation effects. It stabilises the directional 
accuracy of the coherence-enhancing shock filter. A larger $\rho$ usually 
gives a better directional accuracy. Therefore one should usually choose 
$ \rho > \sigma$. The edge scale $\nu$ of the weighting function averages 
structure information locally. Especially in the beginning of the evolution, 
there may not be sufficient unambiguous structural information available 
for the shock term to identify meaningful structures. Presmoothing the gradient 
information with a sufficiently large edge scale allows to assign a suitable 
weighting to the diffusion term and the shock term. 

\medskip

\begin{figure}
\centering
\includegraphics[width=0.5\textwidth]{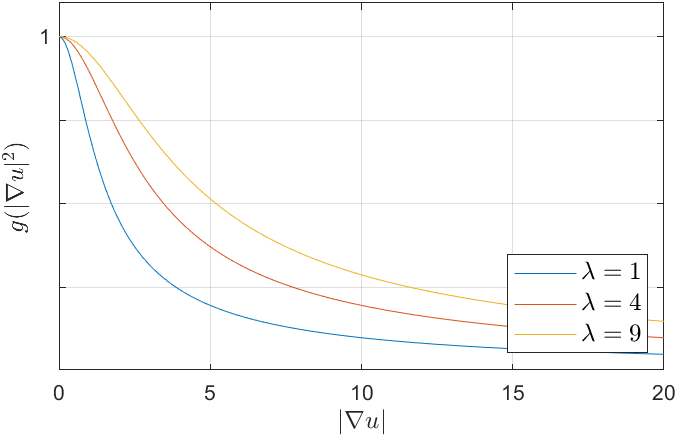}
\caption{Effect of the contrast parameter $\lambda$ 
on the Charbonnier diffusivity.}
\label{fig:charb}
\end{figure}

The contrast parameter $\lambda$ and the regularisation parameter $\varepsilon$
control the tonal behaviour of RDS inpainting in its codomain. As depicted in 
Fig. \ref{fig:charb}, the contrast parameter $\lambda$ determines how fast 
the Charbonnier weight decreases. A smaller $\lambda$ leads to a larger 
zone of gradient values  in which the shock term dominates. The regularisation
parameter $\varepsilon$ softens the transition from dilation to erosion and 
avoids too rapid edge formation in the beginning.
For small values of the second derivative operator in the guidance term,
it shrinks the strength of the shock filter. A small $\varepsilon$ yields
a very harsh transition, and a large $\varepsilon$ results in a more gradual 
evolution towards the discontinuous steady state.

\medskip

In order simplify the parameter optimisation in practice, we calibrate 
the five parameters by a single one in each category. For that purpose, we 
couple the spatial scales to each other. We choose $\rho =\nu= 1.6 \cdot\sigma$
since it works well for all of the experiments that we performed. Moreover, 
we couple the tonal parameters $\lambda$ and $\varepsilon$. In our experiments,
we use $\varepsilon = 0.15 \cdot \lambda $.

\medskip

With that we addressed the main drawback of our conference 
publication~\cite{SW23}: the large number of parameters. 
We reduced the number of parameters that have to be optimised to two. 
This makes RDS inpainting easier to use in practice. Let us also emphasise 
that the newly introduced regularisation allows the parameter coupling 
without a loss in quality in comparison to our original diffusion--shock 
inpainting from \cite{SW23}. Our 
experiments demonstrate that RDS inpainting gives better results than the 
original diffusion--shock inpainting from \cite{SW23} if parameter 
coupling is applied to it.

%..............................................................................
%------------------------------------------------------------------------------
\subsection{Extension to Vector-Valued Data}
Let us now consider a vector-valued image  $f:{\Omega}\to \mathbb{R}^{n_c}$ 
with $n_c$ channels.
Vector-valued data are common in image processing as they are typically used 
for RGB colour images or hyperspectral data. Since our RDS inpainting relies 
on structural information for guidance, a simple 
channelwise application is not appropriate: The shock term might create shocks 
at different locations for each channel, and the weighting
of the shock term and the diffusion term could vary across the channels. 

By utilising a joint squared gradient magnitude~\cite{Di86} as well as a joint 
structure tensor \cite{GKKJ92,We96b}, we can synchronise the operations 
across all channels. The joint Charbonnier weight is given by
\begin{equation}
g\left(\frac{1}{n_c}\sum\limits_{c=1}^{n_c}|\grad (u_c)_\nu|^2\right) \, ,
\end{equation}
and the joint structure tensor is
\begin{equation}
\frac{1}{n_c} \sum\limits_{c=1}^{n_c} \bm J_\rho(\grad (u_c)_\sigma)\, .
\end{equation}
This strategy is also in line with the multi-channel version of the 
coherence-enhancing shock filter from \cite{We03}.

Overall, for the vector-valued  image 
$\bm u:{\Omega}\times [0,\infty) \to \mathbb{R}^{n_c}$ the 
following evolution describes RDS inpainting in the inpainting domain: 
\begin{align}
\label{eq:inpainting-color}
\partial_t u_c \; &= \; 
  g\left(\frac{1}{n_c}\sum\limits_{c=1}^{n_c}|\grad (u_c)_\nu|^2\right)\, 
  \Delta u_c  \nonumber\\
  \;&-\; \left(1-g\left(\frac{1}{n_c}\sum\limits_{c=1}^{n_c}
  |\grad (u_c)_\nu|^2\right)
  \right)\, 
  S_\varepsilon\left( \partial_{\bm w\bm w} ((u_c)_\sigma) 
  \right) \,  | \bm \nabla u_c | 
\end{align}
for each channel $c$. As in the scalar-valued case, we use
Dirichlet data at the boundaries $\partial K$ of the inpainting mask,
and reflecting boundary conditions on the image domain boundary $\partial 
\Omega$.

%%%%%%%%%%%%%%%%%%%%%%%%%%%%%%%%%%%%%%%%%%%%%%%%%%%%%%%%%%%%%%%%%%%%%%%%%%%%%%%
\section{Numerical Algorithm}
\label{sec:discretisation}

In order to apply our model to a discrete image $(f_{i,j})$ with 
pixels $(i,j)$ and grid size $h$, we discretise 
\eqref{eq:inpainting} with an explicit scheme. The discrete evolving image
$u^k_{i,j}$ is an approximation of $u(\bm x,t)$ in the 
cell-centred location $\bm{x}= (i-\frac{1}{2}, j-\frac{1}{2})^\top$ at the
time $t=k\tau$, where $k$ is the iteration number and $\tau$ is the time step
size. For the time derivative, we apply a forward difference
\begin{equation}
 \label{eq:forward}
 (\partial_t u)_{i,j}^k \;=\; \frac{u_{i,j}^{k+1}-u_{i,j}^k}{\tau}.
\end{equation}
The spatial derivatives are evaluated at the old time level $k$. 

%------------------------------------------------------------------------------
\subsection{Approximation of Homogeneous Diffusion}
For the approximation of the homogeneous diffusion term $\Delta u$ we rely 
on the $\delta$-stencil of Welk and Weickert \cite{WW21}. They propose a  
convex combination of axial and diagonal central differences 
in order to achieve a high degree of rotation invariance. The corresponding
stencil is given by
\begin{equation}
\label{eq:dis-diff}
\mbox{
\setlength{\extrarowheight}{0.5mm}
$(\Delta u)_{i,j}^{k}$ $\;=\;$ 
$\Biggl( \dfrac{1-\delta}{h^2}$
\begin{tabular}{|c|c|c|}
\hline
$\;0$ & $\;1$ & $\;0$ \\[0.5mm]
\hline
$\;1$ & $-4$ & $\;1$ \\[0.5mm]
\hline
$\;0$ & $\;1$ & \;$0$ \\[0.5mm]
\hline
\end{tabular}
$ \;+\;  \dfrac{\delta}{2h^2}$
\begin{tabular}{|c|c|c|}
\hline
$\;1$ & $\;0$ & $\;1$ \\[0.5mm]
\hline
$\;0$ & $-4$ & $\;0$ \\[0.5mm]
\hline
$\;1$ & $\;0$ & $\;1$ \\[0.5mm]
\hline
\end{tabular}
$\Biggr) \, u_{i,j}^k\,.$
}
\end{equation}
with a weight $\delta \in [0,1]$. 
{As is common in the numerical literature, the stencil 
notation specifies the discrete convolution weights in the locations}
\vspace{4mm}
\begin{center}
\setlength{\extrarowheight}{1mm}

\begin{tabular}{|c|c|c|}
\hline
$(i\! -\!1, j\! +\!1)$ & $(i, j\!+\!1)$ & $(i\!+\!1, j\!+\!1)$ \\[1mm]\hline
$(i\!-\!1, j)$ & $(i, j)$ & $(i\!+\!1, j)$ \\[1mm]\hline
$(i\!-\!1, j\!-\!1)$ & $(i, j\!-\!1)$ & $(i\!+\!1, j\!-\!1)$ \\[1mm]\hline
\end{tabular}
\end{center}
\vspace{4mm}
An explicit discretisation of the homogeneous diffusion equation 
$\partial_t u = \Delta u$ with this stencil results in the following iterative
scheme:
\begin{align}
u^{k+1}_{i,j} \; &= \; u_{i,j}^k \left(1-  \frac{4-2\delta}{h^2}\tau
                                   \right )
  + \frac{1-\delta}{h^2}\tau\left(  u_{i+1,j}^k 
  +  u_{i-1,j}^k
  +   u_{i,j+1}^k
  +   u_{i,j-1}^k \right ) \nonumber\\
  &\quad + \frac{\delta}{2h^2}\tau \left( u_{i+1,j+1}^k
  +   u_{i+1,j-1}^k
  +   u_{i-1,j-1}^k
  +   u_{i-1,j+1}^k \right)
\end{align}
with $u^0_{i,j} = f_{i,j}$.
Thus, $u^{k+1}_{i,j}$ is a convex combination of the image data at time level 
$k$, if
\begin{equation}
 \label{eq:taud}
 \tau \;\leq\; \frac{h^2}{4-2\delta} \;=:\; \tau_D\, .
\end{equation}
This implies stability in terms of the maximum--minimum principle
\begin{align}
\label{eq:maxmin}
\min_{n,m} f_{n,m} \; \leq\; u^k_{i,j} \;\leq\; \max_{n,m} f_{n,m}\; 
\qquad  \mbox{for all $i$, $j$, and for $k \geq 0$.} 
\end{align}
%---------------------------------------------------------------------------
\subsection{Approximation of Dilation and Erosion}
To discretise the morphological terms $\pm|\bm \nabla u|$, 
{ we rely on upwind schemes. This type of discretisation adaptively
selects a one-sided difference that reflect the local transport direction.
For dilation and erosion, the classical Rouy--Tourin upwind 
schemes~\cite{RT92} are a popular choice. However, for the discretisation of 
the morphological terms $\pm |\bm \nabla u|$ in our RDS inpainting,
we follow Welk and Weickert \cite{WW21} again. They combine the classical 
axial Rouy--Tourin upwind scheme with its diagonal variant with a weight 
$\delta$.} 
For the dilation term $|\bm \nabla u|$, the resulting scheme is given by
\begin{align}
|\bm \nabla u|_{i,j}^k \;=\;
\tfrac{1-\delta}{h} \: \big(
       &\max \, \lbrace u_{i+1,j}^k \!-\! u_{i, j}^k,\;
                     u_{i-1,j}^k \!-\! u_{i, j}^k,\; 0\rbrace^2
                        \nonumber\\
     + &\max \, \lbrace u_{i,j+1}^k \!-\! u_{i, j}^k,\; 
                     u_{i,j-1}^k \!-\! u_{i, j}^k,\;0\rbrace^2
     \big)^\frac{1}{2} \nonumber\\[1mm]
+\;\tfrac{\delta}{\sqrt{2}h} \: \big(
       &\max \, \lbrace u_{i+1,j+1}^k \!-\! u_{i, j}^k,\;
                        u_{i-1,j-1}^k \!-\! u_{i, j}^k,\; 0\rbrace^2
                        \nonumber\\
     + &\max \, \lbrace u_{i-1,j+1}^k \!-\! u_{i, j}^k,\; 
                     u_{i+1,j-1}^k \!-\! u_{i, j}^k,\;0\rbrace^2
      \big)^\frac{1}{2}
\label{eq:dis-dil}     
\end{align}
with the weight $\delta \in [0,1]$ and $u^0_{i,j} = f_{i,j}$.
{The Rouy--Tourin upwind schemes are designed to adapt the one-sided
differences to the local transport direction. Dilation transports bright
values into dark regions, and erosion propagates dark values into bright regions.
Hence, they have opposing transport directions. Therefore, upwind schemes for 
erosion flip the finite differences that are present in the dilation scheme. 
Here, we rely on the work of Welk and Weickert \cite{WW21} as well. They propose to
discretise the erosion term $-|\bm \nabla u|$ as}
\begin{align}
-|\bm \nabla u|_{i,j}^k \;=\;
-\tfrac{1-\delta}{h} \: \big(
       &\max \, \lbrace u_{i,j}^k \!-\! u_{i+1, j}^k,\;
                        u_{i,j}^k \!-\! u_{i-1, j}^k,\; 0\rbrace^2
                        \nonumber\\
     + &\max \, \lbrace u_{i,j}^k \!-\! u_{i, j+1}^k ,\; 
                         u_{i,j}^k \!-\! u_{i, j-1}^k,\;0\rbrace^2
     \big)^\frac{1}{2} \nonumber\\[1mm]
-\;\tfrac{\delta}{\sqrt{2}h} \: \big(
       &\max \, \lbrace u_{i,j}^k \!-\! u_{i+1,j+1}^k,\;
                        u_{i,j}^k \!-\! u_{i-1,j-1}^k,\;0\rbrace^2
                        \nonumber\\
     + &\max \, \lbrace u_{i,j}^k \!-\! u_{i-1,j+1}^k ,\; 
                         u_{i,j}^k \!-\! u_{i+1,j-1}^k ,\;0 \rbrace^2
      \big)^\frac{1}{2}         
      \;.
\label{eq:dis-ero}  
\end{align}
An explicit scheme with forward difference in time
and space discretisation \eqref{eq:dis-dil} or \eqref{eq:dis-ero} results 
in the following iterative schemes for dilation \eqref{eq:iter-dil} and 
erosion \eqref{eq:iter-ero}:
\begin{align}
\label{eq:iter-dil}
u^{k+1}_{i,j} \;=\; u^k_{i,j}+
\tfrac{1-\delta}{h}\tau  \: \big(
       &\max \, \lbrace u_{i+1,j}^k \!-\! u_{i, j}^k,\;
                     u_{i-1,j}^k \!-\! u_{i, j}^k,\; 0\rbrace^2
                        \nonumber\\
     + &\max \, \lbrace u_{i,j+1}^k \!-\! u_{i, j}^k,\; 
                     u_{i,j-1}^k \!-\! u_{i, j}^k,\;0\rbrace^2
     \big)^\frac{1}{2} \nonumber\\[1mm]
+\;\tfrac{\delta}{\sqrt{2}h}\tau \: \big(
       &\max \, \lbrace u_{i+1,j+1}^k \!-\! u_{i, j}^k,\;
                        u_{i-1,j-1}^k \!-\! u_{i, j}^k,\; 0\rbrace^2
                        \nonumber\\
     + &\max \, \lbrace u_{i-1,j+1}^k \!-\! u_{i, j}^k,\; 
                     u_{i+1,j-1}^k \!-\! u_{i, j}^k,\;0\rbrace^2
      \big)^\frac{1}{2},    \\
\label{eq:iter-ero}
u^{k+1}_{i,j} \;=\; u^k_{i,j}
-\tfrac{1-\delta}{h}\tau \: \big(
       &\max \, \lbrace u_{i,j}^k \!-\! u_{i+1, j}^k,\;
                        u_{i,j}^k \!-\! u_{i-1, j}^k,\; 0\rbrace^2
                        \nonumber\\
     + &\max \, \lbrace u_{i,j}^k \!-\! u_{i, j+1}^k ,\; 
                         u_{i,j}^k \!-\! u_{i, j-1}^k,\;0\rbrace^2
     \big)^\frac{1}{2} \nonumber\\[1mm]
-\;\tfrac{\delta}{\sqrt{2}h} \tau\: \big(
       &\max \, \lbrace u_{i,j}^k \!-\! u_{i+1,j+1}^k,\;
                        u_{i,j}^k \!-\! u_{i-1,j-1}^k,\;0\rbrace^2
                        \nonumber\\
     + &\max \, \lbrace u_{i,j}^k \!-\! u_{i-1,j+1}^k ,\; 
                         u_{i,j}^k \!-\! u_{i+1,j-1}^k ,\;0 \rbrace^2
      \big)^\frac{1}{2} \, , 
\end{align}
with $u^0_{i,j} = f_{i,j}$.
They satisfy
the maximum--minimum principle~(\ref{eq:maxmin}) if
\begin{equation}
 \label{eq:taum}
 \tau \;\leq\; \frac{h}{\sqrt{2}\,(1-\delta) + \delta} \;=:\; \tau_M\,.
\end{equation}
In order to show this, one has to consider all possible schemes resulting
from the different cases of the $\max$ operations. For the sake of 
brevity, let us now sketch how to show the stability for the dilation process 
\eqref{eq:iter-dil} in the following case:
\begin{align}
\label{ass1}
\max\lbrace u_{i+1,j}^k\, ,
                u_{i-1,j}^k \, ,
                u_{i,j+1}^k \, ,
                u_{i,j-1}^k \, ,
             u_{i,j}^k 
         \rbrace
\;&=\; u_{i+1,j}^k , \\
\max\lbrace u_{i+1,j+1}^k\, ,
            u_{i-1,j+1}^k \, ,
            u_{i+1,j-1}^k \, ,
            u_{i-1,j-1}^k \, ,
            u_{i,j}^k 
         \rbrace
\;&=\; u_{i+1,j+1}^k \, .
\label{ass2}
\end{align}
Clearly the statement \eqref{eq:maxmin} is fulfilled for $k= 0$ as 
$u^0_{i,j} = f_{i,j}$. Thus, it is sufficient to show that 
$\min\limits_{n,m} u^k_{n,m} \leq u^{k+1}_{n,m} 
\leq \max\limits_{n,m} u^k_{n,m}$.
With \eqref{ass1} and \eqref{ass2}, the dilation scheme \eqref{eq:iter-dil} 
has the upper bound
\begin{align}
u^{k+1}_{i,j} \;&\leq \; 
u_{i,j}^k + \frac{1-\delta}{h}\tau \sqrt{2}(u_{i+1,j}^k - u_{i, j}^k)
+ \frac{\delta}{h}\tau (u_{i+1,j+1}^k - u_{i, j}^k) \nonumber\\
&= \; u_{i,j}^k \left (1- \frac{\sqrt{2}(1-\delta) + \delta}{h} \tau\right ) 
   + \frac{\delta}{h}\tau u_{i+1,j+1}^k 
   + \frac{1-\delta}{h}\tau \sqrt{2}u_{i+1,j}^k
\end{align} 
If \eqref{eq:taum} holds, this is a convex combination. Therefore, we
have 
\begin{align}
u^{k+1}_{i,j}\; \leq \;\max_{n,m} u^k_{n,m}\; \leq \;\max_{n,m} f_{n,m} \, .
\end{align}
Moreover, we have
\begin{align}
u^{k+1}_{i,j} \; \geq \;\min_{n,m} u^k_{n,m} \;\geq\;\min_{n,m} f_{n,m}.
\end{align}
Thus, the discretisation of dilation \eqref{eq:iter-dil} satisfies
a maximum--minimum principle for the case \eqref{ass1}, \eqref{ass2}. The
other cases work in the same way. 
The stability of the erosion evolution can be shown analogously.
 
%------------------------------------------------------------------------------
\subsection{Discretisation of RDS Inpainting}

To discretise the full RDS inpainting equation, we need to discretise the 
guidance term of the shock filter and the weight $g(|\grad u_\nu|^2)$ 
as well. For that we approximate all first order partial derivatives
$\partial_x u$ and $\partial_y u$ in the gradient as well as in the 
structure tensor with Sobel operators \cite{DH73}, since they offer a
good rotation invariance:
\begin{equation}
\mbox{
\setlength{\extrarowheight}{0.5mm}
$(\partial_x u)_{i,j}^k$ $\;\approx\;$ 
$\dfrac{1}{8h}$
\begin{tabular}{|c|c|c|}
\hline
$-1$ & $\;0\;$ & $\;1\;$ \\[0.5mm]
\hline
$-2$ & $\;0\;$ & $\;2\;$ \\[0.5mm]
\hline
$-1$ & $\;0\;$ & $\;1\;$ \\[0.5mm]
\hline
\end{tabular}
 $\;u_{i,j}^k\, ,$

$\quad(\partial_y u)_{i,j}^k$ $\;\approx\;$ 
$\dfrac{1}{8h}$
\begin{tabular}{|c|c|c|}
\hline
$~\,1$ & $~\,2$ & $~\,1$ \\[0.5mm]
\hline
$~\,0$ & $~\,0$ & $~\,0$ \\[0.5mm]
\hline
$-1$ & $-2$ & $-1$ \\[0.5mm]
\hline
\end{tabular}
 $\;u_{i,j}^k$
}.
\end{equation}
The Gaussian convolutions are computed 
in the spatial domain with a sampled and renormalised Gaussian, which is
truncated at five times its standard deviation. We compute the 
normalised dominant eigenvector $\bm w = (c, s)^\top$ of the structure 
tensor analytically, since it is a symmetric $2 \times 2$ matrix. For the 
computation of $\partial_{\bm w\bm w} v$ we use
\begin{equation}
 \left(\partial_{\bm w  \bm w} v\right)_{i,j} ^k
 \;=\; \left(c^2 \, \partial_{xx}v + 2 cs \, \partial_{xy}v + 
                   s^2 \, \partial_{yy}v\right)_{i,j}^k
\end{equation} 
where second order partial derivatives are approximated with the following 
finite differences:
\begin{align}
(\partial_{xx} v)_{i,j}^k  \;&\approx\; 
        \frac{v^k_{i+1,j} - 2v^k_{i,j} + v^k_{i-1,j}}{h^2}\, , \\
(\partial_{yy} v)_{i,j}^k  \;&\approx\; 
        \frac{v^k_{i,j+1} - 2v^k_{i,j} + v^k_{i,j-1}}{h^2}\, ,\\
(\partial_{xy} v)_{i,j}^k  \;&\approx\; 
        \frac{v^k_{i+1,j+1} + v^k_{i-1,j-1} 
              -v^k_{i-1,j+1} - v^k_{i+1,j-1}}{4h^2}\,.
\end{align}
We implement reflecting boundary conditions by adding a layer of mirrored
dummy pixels around the image borders. For the Gaussian convolution of the 
first order derivatives within the structure tensor, we enforce this 
by imposing zero values at the image boundaries.
 
\smallskip

Putting everything together yields the following explicit scheme for 
the RDS inpainting evolution~\eqref{eq:inpainting}:
\begin{align}
\frac{u^{k+1}_{i,j}- u^k_{i,j}}{\tau}\;=\;
  g_{i,j}^k \cdot \big(\Delta u\big)_{i,j}^k - (1-g_{i,j}^k)\cdot 
  S_\varepsilon\left(
  (\partial_{\bm w  \bm w} u_\sigma)^k_{i,j}\right) 
  \, |\bm \nabla u|^k_{i,j}
\label{eq:discrete}
\end{align}
with initial condition $u^0_{i,j} = f_{i,j}$. It inherits its stability
from the schemes for diffusion and morphology:

% This combines a stable discretisation of diffusion with one of morphology. 
% Therefore our scheme inherits this stability if the time step size is bounded
% by $\tau \leq \min\{\tau_M\; , \; \tau_D\} $. For $h \leq 1$ and our chosen 
% $\delta = \sqrt{2}-1$ this results in $\tau_D$.
\medskip

\begin{theorem} {\bf (Stability of the RDS Inpainting Scheme)}\\[1mm]
Let the time step size $\tau$ of the scheme \eqref{eq:discrete}
be restricted by 
\begin{equation}
\tau \; \leq \;  \min\,\{\tau_D, \, \tau_M\}
\end{equation}
with $\tau_D$ and $\tau_M$ as in (\ref{eq:taud}) and (\ref{eq:taum}).\\[1mm]
Then the scheme satisfies the discrete maximum--minimum principle
\begin{align}
\min_{n,m} f_{n,m} \; \leq\; u^k_{i,j} \;\leq\; \max_{n,m} f_{n,m}\;
\qquad  \mbox{for all $i$, $j$, and for $k \geq 0$.}
\end{align}
\end{theorem}

\smallskip
\begin{proof}
If $\tau \leq \min\,\{\tau_D, \, \tau_M\}$, it follows from the stability of 
the diffusion and morphological processes that
\begin{align*}
u^{k+1}_{i,j} \;&=\; u^k_{i,j} +\tau g_{i,j}^k \cdot \big(\Delta u\big)_{i,j}^k
 - (1-g_{i,j}^k)\cdot\,\tau S_\varepsilon
   \big((\partial_{\bm w  \bm w}u_\sigma)^k_{i,j}\big) 
  \, |\bm \nabla u|^k_{i,j}\\
&\leq\; g_{i,j}^k \max_{n,m} f_{n,m} + (1-g_{i,j}^k) 
 \max_{n,m} f_{n,m} \\
 \;&=\;\max_{n,m} f_{n,m} \; .
\end{align*}
Analogously, one can show the condition 
$\, \min\limits_{n,m} f_{n,m} \, \leq \, u^k_{i,j}$.
\end{proof}

For good rotation invariance, we follow the suggestion of Welk and 
Weickert \cite{WW21} and use $\delta =\sqrt{2} - 1$.
Thus, for $\,h=1\,$ our scheme satisfies a maximum--minimum principle for 
$\,\tau \,\leq\, \tau_D \,\approx\, 0.31$. 
This shows a clear advantage of RDS inpainting over EED inpainting 
\cite{SPME14,WW06}, for which there is currently no numerical algorithm that
fulfils a maximum--minimum principle on a bounded stencil. 

In order to use this numerical scheme for vector-valued data, we discretise  
 $\Delta u_c$, $\pm |\bm\nabla u_c|$ and $\partial_t u_c $ for each channel 
 $c \in\{1...n_c\}$
and apply a channel coupling to the weight and structure tensor as indicated 
by Eq. \eqref{eq:inpainting-color}. The stability limit does not change. 
%%%%%%%%%%%%%%%%%%%%%%%%%%%%%%%%%%%%%%%%%%%%%%%%%%%%%%%%%%%%%%%%%%%%%%%%%%%%%%%
\section{Experiments}
\label{sec:experiment}

\subsection{Comparison to Related Approaches}
Combinations of smoothing and shock filtering, either explicitly or implicitly,
are rare in image inpainting, but fairly common for image enhancement. Many 
methods combine mean curvature motion 
(MCM)~\cite{Br78} for smoothing with the shock term of Alvarez and 
Mazorra \cite{AM94}, see e.g. \cite{AM94,Sa96,XPZKYW16}. These methods are
unable to perform inpainting, since MCM is not suitable for inpainting in 
general \cite{CMS98}, and the width of structures propagated by shock 
filters is limited to the presmoothing scale.
Therefore, we compare RDS inpainting with other shock-smoothing 
combinations that rely on homogeneous diffusion instead. This includes 
the methods of Kornprobst et al. \cite{KDA97}, Fu et al.~\cite{FRWC06} and 
Perona--Malik diffusion \cite{PM90}. Table \ref{tb:eq} shows the corresponding 
evolution equations. To isolate the effect of the 
shock term, we also include a variant of RDS inpainting that uses the shock 
term of Alvarez and Mazorra \cite{AM94} instead of the proposed 
coherence-enhancing shock term.

\begin{figure}
\centering
\setlength{\fboxsep}{1pt}
\begin{tabular}{c@{\hspace{2mm}}c@{\hspace{2mm}}c@{\hspace{2mm}}}
\footnotesize{(a) input} & \footnotesize{(b) Perona--Malik} & \footnotesize{(c) Kornprobst et al.}\\
\fbox{\includegraphics[width=0.3\textwidth]{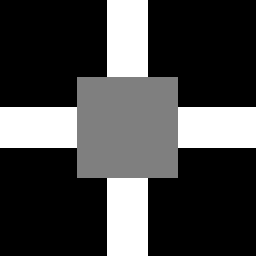}} &
\fbox{\includegraphics[width=0.3\textwidth]{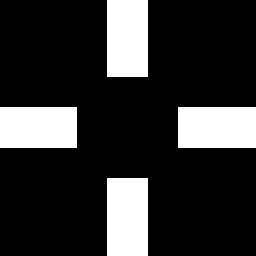}} &
\fbox{\includegraphics[width=0.3\textwidth]{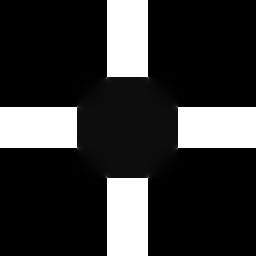}} \\[1mm]
\fbox{\includegraphics[width=0.3\textwidth]{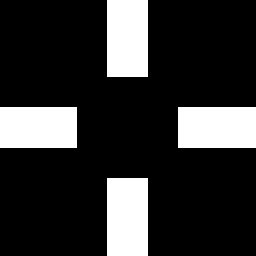}} &
\fbox{\includegraphics[width=0.3\textwidth]{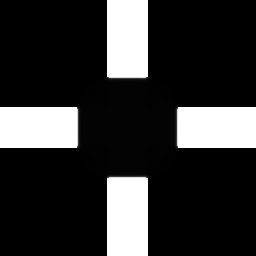}} &
\fbox{\includegraphics[width=0.3\textwidth]{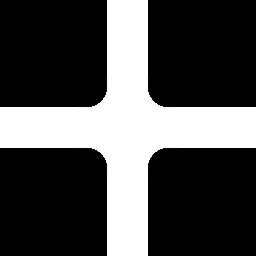}} \\
\footnotesize{(d) Fu et al.} & \footnotesize{(e) AM shock term} & 
\footnotesize{(f) RDS inpainting}
\end{tabular}
\caption{Comparison of RDS inpainting to related approaches. 
Parameters:
(b) $\lambda = 2$. (c) $\beta=\gamma=1$, $\sigma = 1.5$, $T= 30$.
(d) $\ell_1 = 1$, $T_1 = 40$, $T_2 = 35$, $\sigma = 2$. (e)
RDS inpainting with shock term of Alvarez and Mazorra, $\sigma =2$, $\nu = 4$, 
$\lambda = 1.5$. (f) $\sigma =2$, $\lambda = 1.5$. }
\label{fig:related}
\end{figure}

The experiment in Fig. \ref{fig:related} shows the result of our comparison.
It is inspired by a popular experiment for Cahn--Hilliard inpainting 
{from Fig. 2 in the paper \cite{BEG07} }. 
The goal is the reconstruction of a cross. 
Clearly, RDS inpainting gives the best result: It reconstructs a binary, 
cross-like shape. All other methods are unable to connect the white bars. 
Moreover, RDS inpainting also gives a sharper result than the original 
Cahn--Hilliard inpainting { from Fig. 2 in \cite{BEG07}}. 
The comparison 
of  Fig.~\ref{fig:related}~(e) and Fig.~\ref{fig:related}~(f) emphasises 
that the coherence-enhancing shock term is crucial for the performance of 
RDS inpainting. Moreover, it should be noted that our RDS inpainting with 
parameter coupling requires to specify only two parameters, in contrast 
to the competing explicit combinations: The method by Kornprobst et 
al.~\cite{KDA97} uses four parameters, and the approach of Fu et a
l.~\cite{FRWC06} has five different parameters. Thus, our approach is easier 
to use in practice.

%..............................................................................

%------------------------------------------------------------------------------
\subsection{Shape Completion}
Shape completion is a special case of inpainting, in which data is given by a
few parts of a shape. The goal is the reconstruction of the original 
shape. 
This is an especially difficult problem for many inpainting operators: It
requires very high directional accuracy, the ability to bridge large gaps in 
the data and to create perfectly sharp edges. Let us now 
evaluate the performance of our RDS inpainting in the task of shape completion. 

\begin{figure}[tbp]
\centering
\setlength{\fboxsep}{1pt}
\begin{tabular}{c@{\hspace{2mm}}c@{\hspace{2mm}}c}
\fbox{\includegraphics[width=0.25\textwidth]{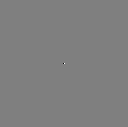}} &
\fbox{\includegraphics[width=0.25\textwidth]{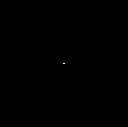}} &
\fbox{\includegraphics[width=0.25\textwidth]{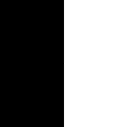}} 
  \\[2mm]
\fbox{\includegraphics[width=0.25\textwidth]{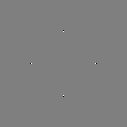}} &
\fbox{\includegraphics[width=0.25\textwidth]{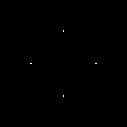}} &
\fbox{\includegraphics[width=0.25\textwidth]{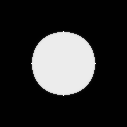}}\\
\footnotesize{(a) input}  & \footnotesize{(b) mask} & 
\footnotesize{(c) RDS inpainting} 
\end{tabular}
\caption{RDS inpainting from dipoles. 
\textbf{Top}: $128\times 128$ image; $\sigma = 2$, $\lambda = 1$.
\textbf{Bottom}:  $127\times 127$ image; $\sigma = 1.8$, $\lambda = 3.2$.}
\label{fig:dipoles}
\end{figure}

Fig. \ref{fig:dipoles} shows two challenging examples. Here the goal is
to reconstruct a half-plane from only one dipole (i.e. a white pixel
next to a black one), and a disk from four dipoles. In both
cases, RDS inpainting shows a flawless performance and recovers 
the desired shapes with the desired sharpness.  

\begin{figure}[tbp]
\centering
\setlength{\fboxsep}{1pt}
\begin{tabular}{c@{\hspace{2mm}}c@{\hspace{2mm}}c@{\hspace{2mm}}c}
\footnotesize{(a) input} &  \footnotesize{(b) hom. diff.} 
& \footnotesize{(c) biharmonic} &  
\footnotesize{(d) TV} \\
{\includegraphics[width=0.23\textwidth]{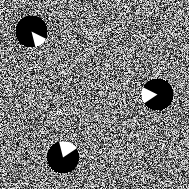}} &
{\includegraphics[width=0.23\textwidth]{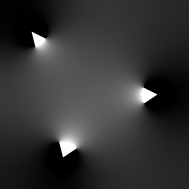}} &
{\includegraphics[width=0.23\textwidth]{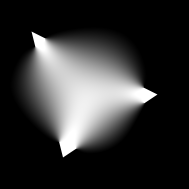}} &
{\includegraphics[width=0.23\textwidth]{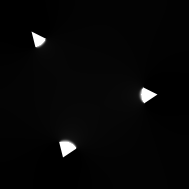}} \\[1mm]
{\includegraphics[width=0.23\textwidth]{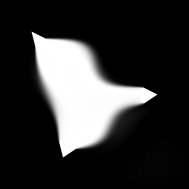}}&
{\includegraphics[width=0.23\textwidth]{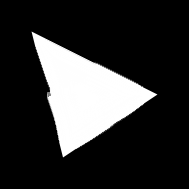}} &
{\includegraphics[width=0.23\textwidth]{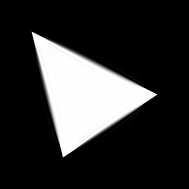}} &
{\includegraphics[width=0.23\textwidth]{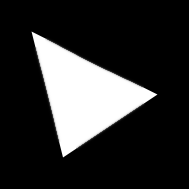}}\\
\footnotesize{(e) Tschumperl\'e} & \footnotesize{(f) BM} & \footnotesize{(g) EED} & \footnotesize{(h) RDS} 
\end{tabular}
\caption{Comparison of inpainting methods. \textbf{Top}:
 Input image with known data in the disks and noise in the unknown region,
 homogeneous diffusion, biharmonic interpolation, and TV inpainting.
\textbf{Bottom}: Tschumperl\'e's approach, Bornemann--M\"arz (BM) method,
 EED inpainting, RDS inpainting with $\sigma = 3.5$ 
 and $\lambda = 3$.
 {All images apart from (h) were provided to us by the authors of 
 \cite{SPME14}}.}
\label{fig:kani}
\end{figure}

To evaluate the performance of RDS inpaiting in comparison to various other 
PDE-based inpainting techniques, we extend an experiment performed by 
Schmaltz et al.~\cite{SPME14} in Fig. \ref{fig:kani}. Inspired by the 
Kanisza triangle, the goal is the reconstruction of a white
triangle on a black background of the data given in the disks. 
Table \ref{tb:eq} shows the energies / evolution equations
associated with each method. Clearly, 
homogeneous diffusion \cite{Ca88} and biharmonic interpolation~\cite{Du76}, 
create a very blurry result which is typical for these linear methods.  
Total variation (TV) inpainting \cite{SC02} fills the whole area in black. 
The directional artefacts created by the method of Tschumperl\'e \cite{Ts06}
hint at suboptimal numerics in the original paper. Due to its suitability 
for connecting level lines, the Bornemann--M\"arz model
\cite{BM07} creates a  satisfactory result but suffers from directional 
inaccuracies. Edge-enhancing diffusion 
(EED)~\cite{SPME14,WW06} reconstructs a perfect triangle. Schmaltz
et al.~\cite{SPME14} attribute this high performance to the anisotropy 
and the semilocality of the approach. By semilocality they refer to the 
fact, that EED uses neighbourhood information rather than acting
purely local. RDS inpainting shares these properties. The coherence-enhancing
shock term introduces a strong anisotropy, and the presmoothing procedures
create semilocality. It also creates a high quality result without any 
directional artefacts, and the created edges are even sharper than those
created by EED. Additionally, in contrast to EED our numerical algorithm 
for RDS inpainting also provides a maximum--minimum principle in the discrete
case.

\medskip

\begin{figure}[tb]
\centering
\setlength{\fboxsep}{1pt}
\begin{tabular}
{c@{\hspace{2mm}}c@{\hspace{2mm}}c@{\hspace{2mm}}c@{\hspace{2mm}}c}
\includegraphics[width=0.178\textwidth]{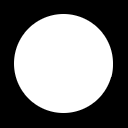} &
\includegraphics[width=0.178\textwidth]{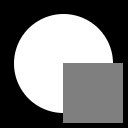} &
\includegraphics[width=0.178\textwidth]{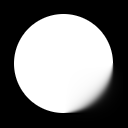} &
\includegraphics[width=0.178\textwidth]{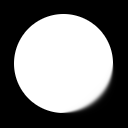} &
\includegraphics[width=0.178\textwidth]{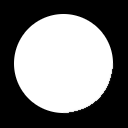} \\[1mm]
\fbox{\includegraphics[width=0.17\textwidth]{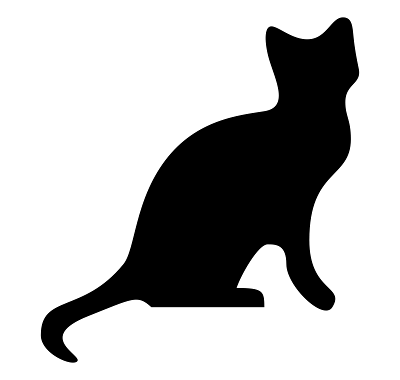}} &
\fbox{\includegraphics[width=0.17\textwidth]{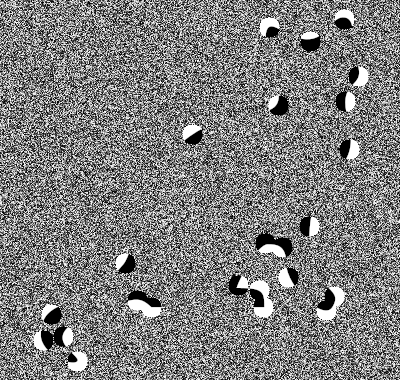}} &
\fbox{\includegraphics[width=0.17\textwidth]{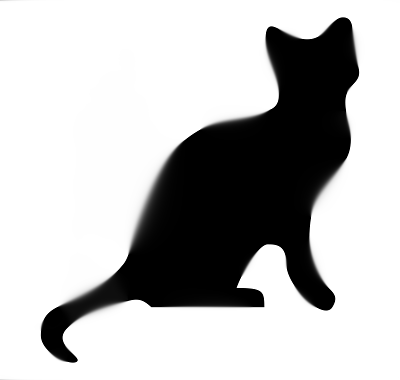}} &
\fbox{\includegraphics[width=0.17\textwidth]{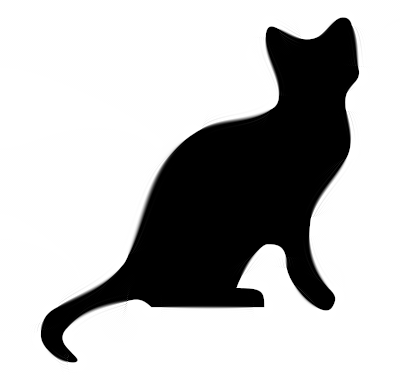}} &
\fbox{\includegraphics[width=0.17\textwidth]{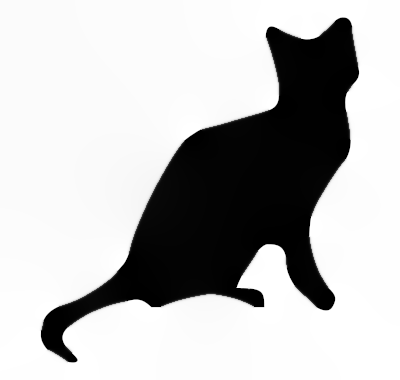}} \\
\footnotesize{(a) original} & \footnotesize{(b) input} & \footnotesize{(c) elastica} & 
\footnotesize{(d) EED} & \footnotesize{(e) RDS} 
\end{tabular}
\caption{Comparison of Euler's elastica, EED, and RDS inpainting.
Parameters for RDS inpainting:
\textbf{Top}: $\sigma = 2.5$ and $\lambda = 2$.
\textbf{Bottom:}
$\sigma = 2.1$ and $\lambda = 5.5$.}
\label{fig:cat}
\end{figure}

In Fig. \ref{fig:cat}, we compare the performance of RDS inpainting to 
EED~\cite{We94e} and Euler's elastica~\cite{MM98a,Mu94a}, two methods that 
produce state-of-the-art results in the context of shape completion. Table
\ref{tb:eq} shows the evolution equation of EED and the energy functional 
that corresponds to Euler's elastica. The results of Euler's elastica are 
published in \cite{SAWE22}  { and were given to us by the authors}. 
The cat data and the EED inpainting of the cat {are the original 
images from \cite{We12} that were provided to us by the author}. 
In both examples RDS inpainting shows similar 
results as EED and Euler's elastica. Overall, RDS inpainting creates
the sharpest results.

%.............................................................................

%..............................................................................

%.............................................................................
\begin{table}
\centering
\caption{PDEs or energies corresponding to the inpainting operators
that we compare to in our experiments. We use the following notations:  
$\bm D $: Diffusion tensor,
$\mathcal{H}$: Hessian, 
$\bm a_\phi = (\cos \phi, \sin\phi)^\top$, 
$\bm{\mathcal{{J}}}_{\sqrt{\bm T}\bm a_\phi}$:
 Jacobian of the vector field $\Omega\to \sqrt{\bm T}\bm a_\phi$, 
$\bm T$: smoothing tensor,
$\alpha, \beta, \gamma, \ell_1, \ell_2, b$: weights, 
$T_1, T_2, T$: thresholds,
$\kappa$: curvature,
$\mbox{th}$: $\tanh$,
$\mbox{tr}$: trace.
 }
\label{tb:eq}
\begin{tabular}{|c|c|}
\hline &\\
Operator & Evolution Equation / Energy Functional\\[2mm]
\hline &\\
Hom. Diff.~{\cite{Ii62,Ii63a,WII97}} & 
  $\partial_t u \;=\; \Delta u$
\\[5mm]
Biharm. Interpol.~\cite{Du76} & $\partial_t u \;=\; -\Delta^2 u$
\\[5mm]
Perona--Malik \cite{PM90} & 
  $\partial_t u \;=\; \bm{\mbox{div}}
    \left(\frac{\bm\nabla u}{1 + |\bm\nabla u|^2/\lambda^2}\right)$
\\[5mm]
Kornprobst et al. \cite{KDA97}& 
  $\partial_t u \;=\; 
  \begin{cases}
  \beta\Delta u & \mbox{for } T < |\bm\nabla u_\sigma|,\\
  \beta\Delta u  - 
  \gamma \sgn(\partial_{\bm\eta\bm\eta} u_\sigma) |\bm\nabla u|
  				& \mbox{for }  T \geq |\bm\nabla u_\sigma|.
  \end{cases}  
  $
\\[7mm]
Fu et al. \cite{FRWC06}&
\resizebox{9cm}{!}{
$\partial_t u \, = \,
  \begin{cases}
  \frac{\partial_{\bm{\xi\xi}} u}{1+\ell_1 \partial_{\bm{\xi\xi}} u}
  - \mbox{\small sgn}(\partial_{\bm{\eta\eta}} u_\sigma)|\bm\nabla u|      
  &\mbox{for } |\bm\nabla u_\sigma| >T_1,\\
  
  \Big(
  \frac{\partial_{\bm{\xi\xi}} u}{1+\ell_1 \partial_{\bm{\xi\xi}} u}
  - |\mbox{\small th}(\ell_2 \partial_{\bm\eta\bm\eta} u)
    |\mbox{\small sgn}(\partial_{\bm\eta\bm\eta} u_\sigma)|\bm\nabla u|
 \Big)
   &  \mbox{for } |\bm\nabla u_\sigma| \in (T_2, T_1],\\
 \Delta u								& \mbox{else}.
  \end{cases}$
}
\\[8mm]
Total Variation \cite{SC02}  & 
  $E(u) \;=\; \int\limits_{\Omega} 
  \Big(\frac{1}{2}(u-f)^2 - \alpha |\bm\nabla u|\Big) d\bm x$
\\[5mm]
Tschumperl\'e \cite{Ts06} &  
$\partial_t u = \mbox{tr}(\bm T \mathcal{H}) 
  + \frac{2}{\pi} (\bm\nabla u)^\top 
  \int\limits^\pi_{0} \bm{\mathcal{{J}}}_{\sqrt{\bm T}\bm a_\phi}
  \sqrt{\bm T} \bm a_\phi d\phi
$
\\[5mm]
Bornemann--M\"arz~\cite{BM07}&  algorithmic approach\\[5mm]
EED~\cite{We94e} & $\partial_t u \;=\; \bm{\mbox{div}}
    \left(\bm D \bm \nabla u\right)$ \\[5mm]
Euler's Elastica~\cite{MM98a,Mu94a} &
$E(u) \;=\; \int\limits_{\Omega} 
		|\bm\nabla u|(b+(1-b)\kappa^2 (u)) d\bm x$  
\\[5mm]
\hline
\end{tabular}
\end{table}

%------------------------------------------------------------------------------
\subsection{Evaluation of the Guidance Function}
\begin{figure}[tbp]
\centering
\setlength{\fboxsep}{1pt}
\begin{tabular}{c@{\hspace{2mm}}c@{\hspace{2mm}}c@{\hspace{2mm}}}
{\includegraphics[width=0.3\textwidth]{img/kani/kani}} &
{\includegraphics[width=0.3\textwidth]{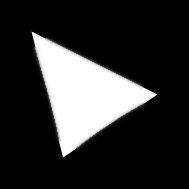}}&
{\includegraphics[width=0.3\textwidth]{img/kani/kani-ds-atan}}\\
\footnotesize{(a) input} &  
\footnotesize{(b) $S_0(x) = \sgn(x)$} 
& \footnotesize{(c) 
$S_\varepsilon = \frac{2}{\pi}\arctan(\frac{x}{\varepsilon})$} \\
\end{tabular}
\caption{Comparison of diffusion--shock inpainting with parameter coupling 
guided by a $\sgn$ and $\arctan$ function.
Parameters: (b) $\sigma =5.8$, $\nu=\rho=1.8\, \sigma$, $\lambda = 3.5$, and 
(c) $\sigma = 3.5$, $\lambda = 3$.}
\label{fig:kani-sgn}
\end{figure}

In the model of RDS inpainting, we use an $\arctan$ function as 
the guidance function, whereas the original diffusion-shock inpainting 
from~\cite{SW23} relies on the $\sgn$ function. While the
original model has already provided high quality results, it requires
the optimisation of four parameters, which makes the method difficult to 
use in practice. 
In order to address this drawback, we have proposed a parameter coupling in a 
previous section. Applying these ideas to diffusion--shock inpainting based on 
a $\sgn$ function decreases the inpainting quality. 
This is not the case for RDS inpainting. Fig.~\ref{fig:kani-sgn} 
demonstrates this by the triangle reconstruction example. Clearly, the 
$\arctan$-guided result is superior: It creates sharper edges and reproduces a 
better directional accuracy. 

\begin{figure}[tbp]
\centering
\setlength{\fboxsep}{0pt}
\begin{tabular}
{c@{\hspace{2mm}}c@{\hspace{2mm}}c@{\hspace{2mm}}c@{\hspace{2mm}}}
\fbox{\includegraphics[width=0.23\textwidth]{img/cat/cat-data}} &
\fbox{\includegraphics[width=0.23\textwidth]{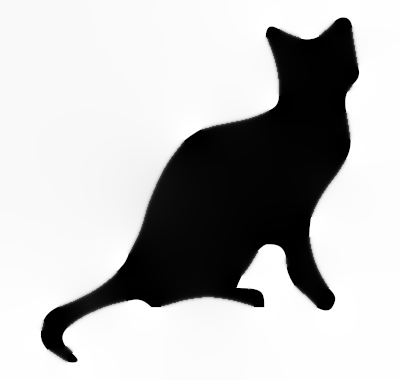}}&
\fbox{\includegraphics[width=0.23\textwidth]{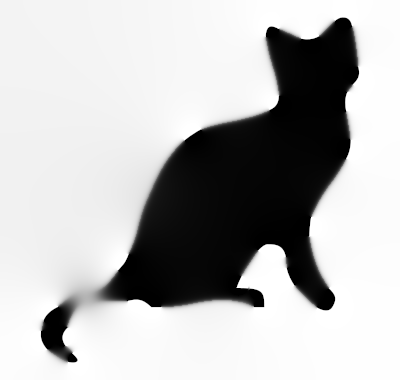}}&
\fbox{\includegraphics[width=0.23\textwidth]{img/cat/cat-ds-atan}}\\
\footnotesize{(a) input} &  
\makecell{\footnotesize{(b) $S_0(x) = \sgn(x)$},\\ 
  \footnotesize{$4$ parameters}}
& \makecell{\footnotesize{(c) $S_0(x) = \sgn(x)$},\\ 
  \footnotesize{$2$ coupled parameters}}
& \makecell{\footnotesize{(d) 
     $S_\varepsilon = \frac{2}{\pi}\arctan(\frac{x}{\varepsilon})$}\\
  \footnotesize{$2$ coupled parameters}
  } \\
\end{tabular}
\caption{Comparison of diffusion--shock inpainting with parameter coupling 
guided by a $\sgn$ and $\arctan$ function, and the result from our conference
publication.
Parameters: (b) $\sigma = 4.2$, $\rho=4.8$, $\nu=4.5$, and $\lambda = 7$, 
(c) $\sigma = 4.3$, $\lambda = 5.4$, $m= 1.8$ and (d) $\sigma = 2.1$ and 
$\lambda = 5.5$.}
\label{fig:cat-sgn}
\end{figure}

{
The cat reconstruction experiment in Fig.~\ref{fig:cat-sgn} makes this 
even more apparent. Here, we compare 
the result from Fig. 5 of our conference publication~\cite{SW23} which used 
a $\sgn$-guided diffusion-shock inpainting without parameter coupling (b),
$\sgn$-guided diffusion-shock inpainting with parameter coupling (c) and 
RDS inpainting with parameter coupling (d). Clearly, RDS inpainting with 
parameter coupling creates a result that is very similar to the conference 
publication. However, the $\sgn$-guided diffusion-shock inpainting is not 
able reconstruct the cat in a satisfactory way. This highlights the necessity 
of the regularisation in RDS inpainting for a parameter coupling that does 
not lead to a loss of inpainting quality.} 

%..............................................................................

%------------------------------------------------------------------------------
\subsection{Natural Images}
\begin{figure}
\centering
\begin{tabular}{c@{\hspace{2mm}}c@{\hspace{2mm}}c@{\hspace{2mm} }}
{\includegraphics[width=0.3\textwidth]{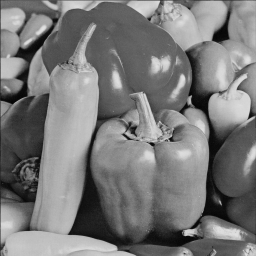}} &
{\includegraphics[width=0.3\textwidth]{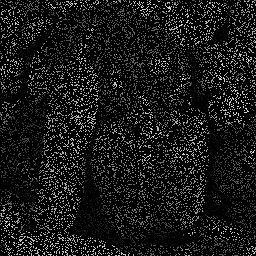}} &
{\includegraphics[width=0.3\textwidth]{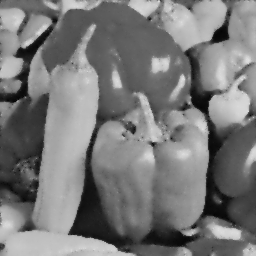}} \\[1mm]
{\includegraphics[width=0.3\textwidth]{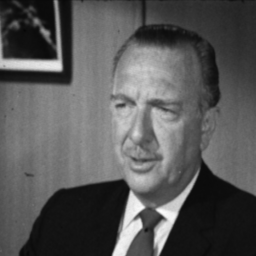}} &
{\includegraphics[width=0.3\textwidth]{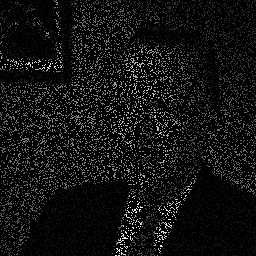}} &
{\includegraphics[width=0.3\textwidth]{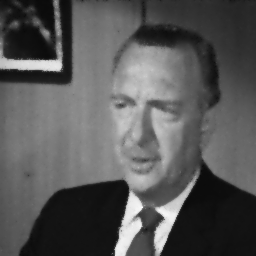}} \\[1mm]
{\includegraphics[width=0.3\textwidth]{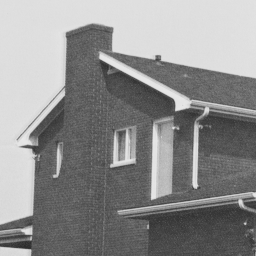}} &
{\includegraphics[width=0.3\textwidth]{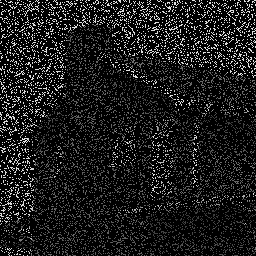}} &
{\includegraphics[width=0.3\textwidth]{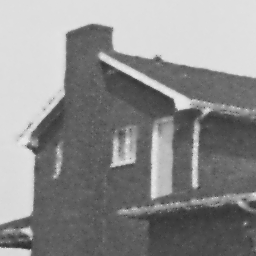}} \\[1mm]
\footnotesize{ (a) original, $256\times 256$} &  \footnotesize{(b) input} & 
\footnotesize{(c) RDS inpainting} \\
\end{tabular}
\caption{RDS inpainting of sparse greyscale images ($20$\% randomly selected 
         pixels). \textbf{First row}: $\sigma = 1.5$, $\lambda = 5$. 
         \textbf{Second row}: $\sigma = 2.1$, $\lambda = 4$. 
         \textbf{Third row}:
         $\sigma = 2.1$, $\lambda = 4.5$.}
\label{fig:grey}
\end{figure}

\begin{figure}[tbp]
\centering
\begin{tabular}{c@{\hspace{2mm}}c@{\hspace{2mm}}c@{\hspace{2mm}}}
{\includegraphics[width=0.3\textwidth]{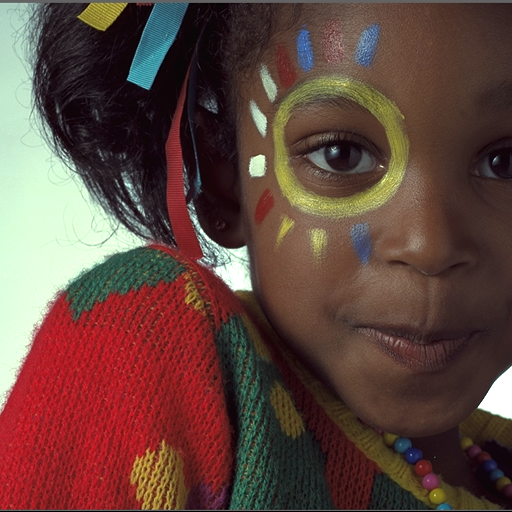}} &
{\includegraphics[width=0.3\textwidth]{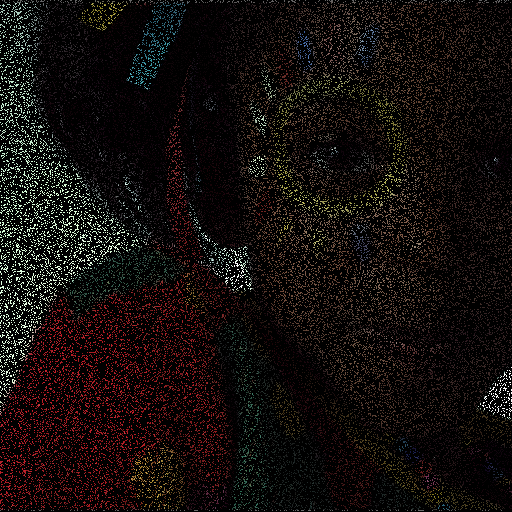}} &
{\includegraphics[width=0.3\textwidth]{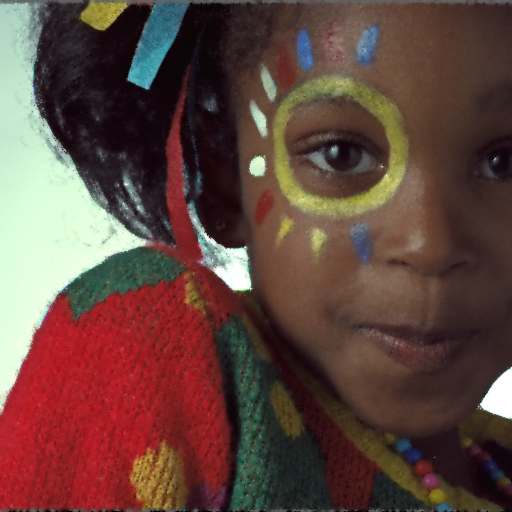}} \\[1mm]
{\includegraphics[width=0.3\textwidth]{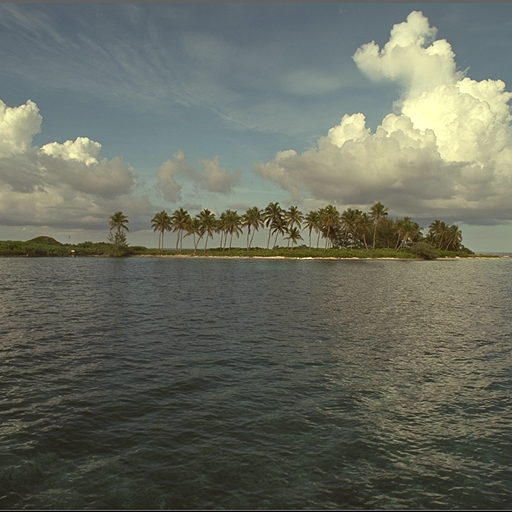}} &
{\includegraphics[width=0.3\textwidth]{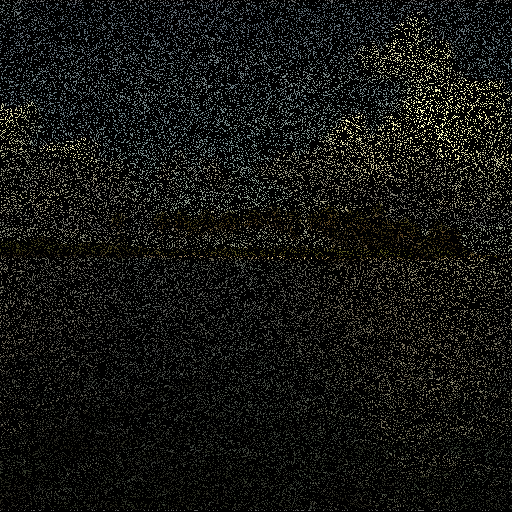}} &
{\includegraphics[width=0.3\textwidth]{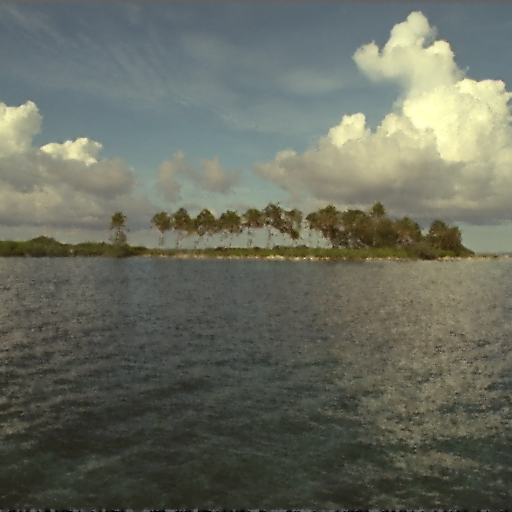}} \\[1mm]
{\includegraphics[width=0.3\textwidth]{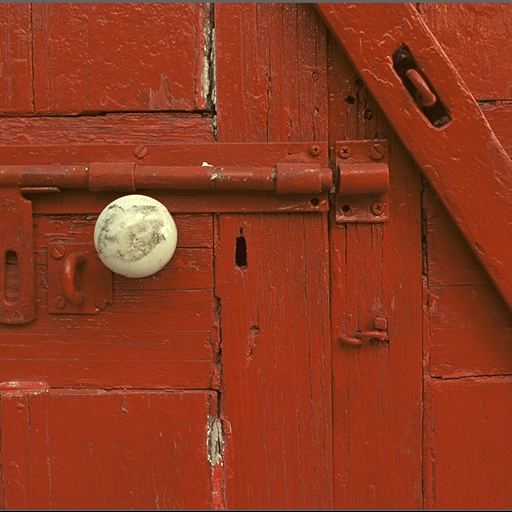}} &
{\includegraphics[width=0.3\textwidth]{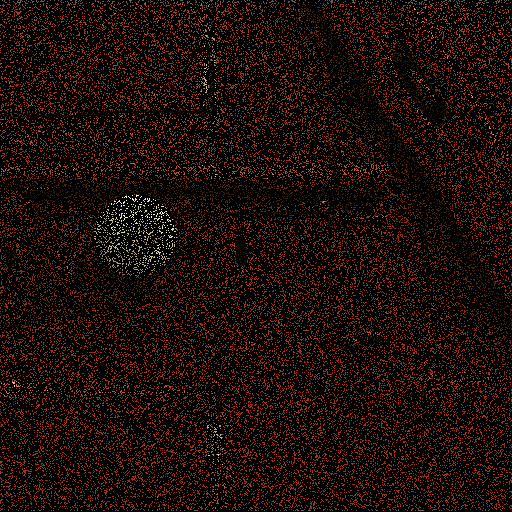}} &
{\includegraphics[width=0.3\textwidth]{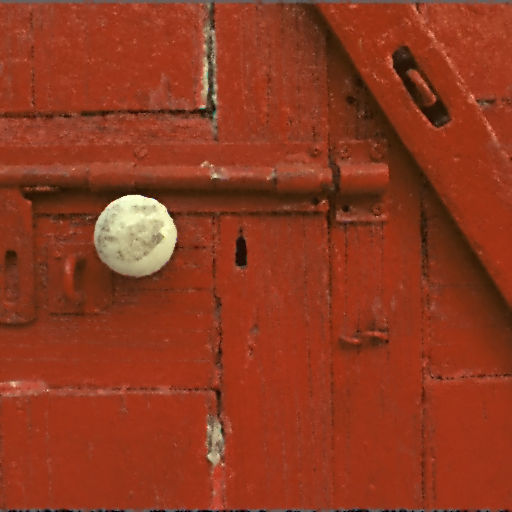}} \\
\footnotesize{ (a) original, $512\times 512$} &  \footnotesize{(b) input} & 
\footnotesize{(c) RDS inpainting} \\
\end{tabular}
\caption{RDS inpainting for sparse colour images ($20$\% randomly selected 
         pixels).
         \textbf{First row}: $\sigma = 3$, $\lambda = 3$.
         \textbf{Second row}: $\sigma = 1.5$, $\lambda = 5$.
         \textbf{Third row}:  and $\sigma = 1.5$, $\lambda = 4$.
         }
\label{fig:colour}
\end{figure}

So far, we considered only binary images since they are especially challenging
for PDE-based inpainting techniques. In Fig.~\ref{fig:grey} and 
Fig.~\ref{fig:colour}, we show that RDS inpainting is also a 
suitable method for 
the reconstruction of natural images from sparse data. Fig.~\ref{fig:grey} 
shows this for greyscale images of size $256 \times 256$. There, the  
runtime was 
% around $3.5$ seconds.
$2.3$ seconds for the {\em peppers} image, $2.5$ seconds for the {\em walter} 
image and $2$ seconds for the {\em house} image on a PC with an 
Intel\textsuperscript{\textcopyright} Core\texttrademark i9-11900K CPU @ 
3.50 GHz.

Fig. \ref{fig:colour} depicts several RDS inpainting results created from
sparse colour images. The original images are cropped versions of images 
from the Kodak dataset \cite{Ea99}. The sparse data are created by randomly 
selecting 20\% of the 
pixels. The results show the effect of using a joint structure tensor and 
a joint weighting function for vector-valued images: Edges are formed in a 
synchronised way, and no unexpected colours or 
colour artefacts are introduced. 

%..............................................................................
 
%..............................................................................
%%%%%%%%%%%%%%%%%%%%%%%%%%%%%%%%%%%%%%%%%%%%%%%%%%%%%%%%%%%%%%%%%%%%%%%%%%%%%%%
\section{Conclusions and Future Work}
\label{sec:conclusions}
We have proposed regularised diffusion--shock (RDS) inpainting  as an extension 
of our diffusion--shock inpainting from \cite{SW23}. Diffusion--shock 
inpainting is the first method to utilise the perfect sharpness and directional
accuracy of a coherence-enhancing shock filter \cite{We03}  in the field of 
inpainting. Together with homogeneous diffusion 
{\cite{Ii62,Ii63a,WII97}}, it 
creates results that rival the quality of  popular PDE-based inpainting 
operators such as edge-enhancing diffusion~\cite{We94e} and Euler's 
elastica~\cite{MM98a,Mu94a}. However, in contrast to these methods, its 
numerical algorithm also satisfies a maximum--minimum principle in the discrete 
case.
\medskip

RDS inpainting introduces a regularisation to the original model. It stabilises 
the model w.r.t. the choice of parameters, and 
thereby allows the reduction of the number of parameters to two. 
This solves the largest disadvantage of the original diffusion--shock inpainting 
model from \cite{SW23}.

\medskip

RDS inpainting is a second order integrodifferential process consisting of two
simple components: homogeneous diffusion and coherence-enhancing shock 
filtering. We showed that it can offer equal or higher quality than higher order 
methods. However, higher order methods are algorithmically more challenging and 
often do not provide stability guarantees. On the other hand, our RDS inpainting 
allows a simple discretisation with an explicit scheme that 
provides a maximum--minimum principle. It constitutes a 
high quality second order integrodifferential process that 
questions the necessity of higher order methods in practice. 
This highlights the potential behind this class of methods, and we
are aiming at gaining a deeper understanding of 
such integrodifferential processes in our ongoing work.

\smallskip
Most PDEs for inpainting are elliptic or parabolic. However, our results 
emphasise that hyperbolic processes deserve far more attention. They
are a natural concept for modelling discontinuities, and shock 
filters are a prototype for this. For our application the coherence-enhancing
shock filter in combination with homogeneous diffusion is the ideal choice. 
Interestingly, both components have been around for at least 20 years. 
This indicates that 
there still lies a huge potential in PDE-based inpainting, especially in 
hyperbolic concepts. Thus, we aim at exploring them further in our future 
work.

\subsubsection*{Acknowledgements.} We thank Karl Schrader for providing 
us with the images and results from his publication \cite{SAWE22}.\\
This project has received funding from the European Research Council
(ERC) under the European Union's Horizon 2020 research and innovation
programme (grant agreement No. 741215, ERC Advanced Grant INCOVID).

\bibliography{refs.bib}

\end{document}